\documentclass[12pt]{article}
\usepackage{latexsym}\usepackage{amsmath}\usepackage{amssymb}
\begin{document}
\markboth{}{Non-relativistic limit of the Einstein equation}
\newcommand{\be}{\begin{equation}}
\newcommand{\ee}{\end{equation}}
\newcommand{\sprd}[2]
{\left\langle#1\,,#2\right\rangle\,}
\newcommand{\prt}{\partial}
\newcommand{\pder}[2]{\frac{\partial #1}{\partial #2}}
\newcommand{\half}{\frac{1}{2}}
\newcommand{\quart}{\frac{1}{4}}
\def \con {\mbox{const}}
\begin{center}{\large\bf Non-relativistic limit of the Einstein
equation}\\[5mm] Turakulov Z. Ya.\\ {\it  Ulugh Bek Astronomy
Institute (UBAI),\\ Astronomicheskaya 33, Tashkent 700052,
Uzbekistan\\ and Institute for Advanced Studies in Basic Sciences
(IASBS)\\ P. O. Box 45195-1159 Zanjan 45195 Iran}\end{center}
\begin{abstract} In particular cases of stationary and stationary
axially symmetric space-time passage to non-relativistic limit of
Einstein equation is completed. For this end the notions of absolute
space and absolute time are introduced due to stationarity of the
space-time under consideration. In this construction absolute time
is defined as a function $t$ on the space-time such that $\prt_t$ is
exactly the Killing vector and the space at different moments is
presented by the surfaces $t=\con $. The space-time metric is
expressed in terms of metric of the 3-space and two potentials one
of which is exactly Newtonian gravitational potential $\Phi$,
another is vector potential $\vec A$ which, however, differs from
vector potential known in classical electrodynamics. In the
first-order approximation on $\Phi/c^2$, $|\vec A|/c$ Einstein
equation is reduced to a system for these functions in which
left-hand sides contain Laplacian of the Newtonian potential,
derivatives of the vector potential and curvature of the space and
the right-hand sides do 3-dimensional stress tensor and densities of
mass and energy. Subj-class: Classical Physics. Keywords: general
relativity; non-relativistic limit; non-flat space.\end{abstract}\section{Introduction}

Newtonian theory of gravitation was used as the underlying base when
constructing general relativity and remains a good non-relativistic
approximation to it. However, prediction of the Lense-Thirring
effect few years after it was created gave rise to the idea of
Coriolis or gravimagnetic interaction together with the
corresponding field which requires some extension of the Newtonian
theory. Discussing a question which could well be answered centuries
ago would show that the Newtonian theory is incomplete. Indeed,
consider a spherical mass of self-gravitating matter in equilibrium
without rotation. The  same equilibrium state should exist in
rotating frame. However, passage to a frame rotating with angular
velocity $\omega$ produces in cylindric coordinates the centrifugal
potential $\omega^2\rho^2/2$ while rotation of the matter in this
frame produces exactly the same centrifugal potential. Instead of
canceling each other, these two add that breaks the equilibrium.
Therefore equilibrium of non-rotating matter in a rotating frame
cannot be described in Newtonian theory. However, {\it The laws of
physics must be of such a nature that they apply to systems of
reference in any kind of motion} \cite{AE1916}. Inclusion of
gravimagnetic field whose strength in this example equals
$H_z=\omega^2\rho$ improves the situation because velocity of matter
multiplied by this strength cancels both the centrifugal potentials.

More recent development of astrophysics brought new questions to be
answered. One of them consists in the following. Infinitesimally-thin
disk is an important astrophysical model. Its mass density is singular
in space that can be presented as one-dimensional $\delta$-function
multiplied by surface density. Due to Einstein equation curvature of
the space-time has the same singularity, hence, the factor $c^{-2}$
where $c$ is speed of light, which the curvature is multiplied by,
does not provide flatness assumed in the Newtonian theory. In other
words, passage to ``Newtonian'' limit does not lead to Newtonian
theory. This fact signifies an apparent need for the next
post-Newtonian approximation in which $c^{-2}$ terms are included and
the space is not assumed to be flat.

Another reason to build this approximation is need for models of
neutron stars in astrophysics. Neutron stars are known to possess
the highest possible mass density in the nature, which is about
$10^{14}g\cdot cm^{-3}$. At the same time velocity of matter in
these objects does not exceed $0.1\ c$, therefore, non-relativistic
mechanics provides sufficient accuracy if applied under these
conditions. Gravitational potential measured in units of $c^2$ does
not exceed the value of $10^{-2}$ \cite{kalobaym}, thus, is usable.
Then equilibrium of matter might well be described in terms of
Newtonian or post-Newtonian approximation which assumes that time is
absolute, space is flat and absolute and gravitational field is
presented by two potentials: one scalar (gravistatic) and one vector
(gravimagnetic, which produces the Coriolis force
\cite{AE1913,Browne}). However, the absolute space cannot be put
flat as in these approximations. The mass density multiplied by
gravitational constant which is $1.87\cdot10{-27}\ cm\ g^{-1}$
yields typical value of curvature $3.9\cdot10^{-13}cm^{-2}$. Though
this value seems to be negligibly small, employing typical radius of
neutron star which is about $10^6 cm$ as the unit of length leads to
spatial curvature about $0.39$ that cannot be neglected.

Consequently, the most appropriate approximation to general
relativity to be used for this end would be a non-relativistic
theory with absolute space and time and gravitational field
specified by the potentials and geometry of the space. Generally
accepted equations of this theory are $$\square h^{\alpha\beta}=
-16\pi\tau^{\alpha\beta},$$ where $\square\equiv\partial^2/\partial
t^2+\nabla^2$ is the `flat-space-time wave operator',
$h^{\alpha\beta}$ is a ``gravitational tensor potential'' related to
the deviation of the space-time metric from its Minkowski form by
the formula $h^{\alpha\beta}\equiv \eta^{\alpha\beta}- (-g)^{-1/2}
g^{\alpha\beta}$, g is the determinant of $g_{\alpha\beta}$
\cite{clwill}. In this work another approximation to Einstein
equation is considered, in which all velocities are assumed to be
negligibly small compared with $c$, therefore time and space are
absolute, but the space is not assumed to be flat. The goal of the
present work is to derive exact consequences from the Einstein
equation under these assumptions.

In order to determine what will be called ``space'' and ``time'' we
introduce a coordinate system $\{t,x^i\}$ in which the coordinate $t$
plays the key role because thereafter this coordinate will be used as
absolute time and surfaces $t=const$ -- absolute space. As a function
on the space-time this coordinate is to be chosen such a way that
square of its gradient $\sprd{dt}{dt}$ is close enough to unit due to
the approximation chosen. Afterwards, whatever coordinate
transformations we make, the coordinate $t$ remains unchanged. It must
be noted that $t$ is only a coordinate, not proper time of an
observer because length of its gradient $dt$ is not exactly one.

By spatial metric we mean Riemannian metric of the surfaces
$t=const$ $$\sprd{dx^i}{dx^j}|_{t=const}\equiv g^{ij}.$$ Since
functions $x^i$ on the space-time are quite arbitrary, genuine
gradients $dx^i$ are not orthogonal to $dt$: $\sprd{dt}{dx^i}\equiv
A^i/c$ and genuine scalar products $\sprd{dx^i}{dx^j}$ differ from
$g^{ij}$, however the difference is of order $c^{-2}$ and we neglect
it. By the result we have space-time metric in the form
\be\sprd{dt}{dt}=1-\frac{2\Phi}{c^2},\, \sprd{dt}{dx^i}=
\frac{A^i}{c},\, \sprd{dx^i}{dx^j}=-g^{ij}-\frac{A^iA^j}{c^2}.
\label{gmetric}\ee It will be shown below that the term
$\Phi/c^2$ in $g^{tt}$ cannot be neglected, as well as
$g^{t\varphi}$ which is of order $c^{-1}$ while for purely spatial
components they are to be omitted.\section{Coordinate transformations}

Intersection of three coordinate surfaces $x^i=const$ is coordinate
line, a time-like curve which specifies the time axis through given
point of the space. In general, the time axis and the space are not
orthogonal because, on one hand, this curve is specified only by
coordinate surfaces $x^i=const$ and, on the other hand, these
surfaces are introduced regardless of choice of the surfaces
$t=const$. Non-orthogonality of space and time reveals in the
components $g^{ti}$ of the metric (\ref{gmetric}). In this section
we consider coordinate transformations under which the space and
time remain unchanged.

Invariance of the space means that the transformations do not touch
the coordinate $t$, thus, they are of the form$$t\rightarrow t,\
x^i\rightarrow y^a(x^i,t).$$ These transformations are of two
different kinds. Ordinarily coordinate transformations$$t\rightarrow
t,\ x^i\rightarrow y^a(x^i)$$ leave the time axis immobile
everywhere. Another kind is given by changes of frame of reference.
Its simplified form is\be x^i\rightarrow y^i=x^i+f^i(t)
\label{moframe}\ee and the general form is superposition of
transformations of both kinds. Hereafter we employ only simplified
form (\ref{moframe}) of this transformation.

Differentiation of the equation (\ref{moframe}) gives$$ dy^i=dx^i+
f^{\prime i}dt.$$ It is seen that the metric (\ref{gmetric}) remains
invariant under change of frame, proviso that coefficients $A^i$
transform as\be A^i\rightarrow A^i+cf^{\prime i}\left(
1-\frac{2\Phi}{c^2}\right).\label{Aito} \ee\section{Hamilton-Jacobi
equation}

Hamilton-Jacobi equation for time-like geodesics has the form\be
\left(1-\frac{2\Phi}{c^2}\right)\left(\pder{S}{t}\right)^2+
\frac{2A^i}{c}\,\pder{S}{t}\pder{S}{x^i}-\sprd{\nabla S}{\nabla
S}=m^2c^2 \label{HJEgen}\ee where $\nabla S$ stands for the spatial
part of the 1-form $dS$. Assume that there exists a function $W$
satisfying the equations\be 2m\pder{W}{t}=
\left(\pder{S}{t}\right)^2-m^2c^2,\ \nabla S=\nabla W\label{shorten}
\ee (for example, $S_t$ depends only on $t$). Then the equation
(\ref{HJEgen}) can be rewritten as follows:
\begin{eqnarray*}
\left(1-\frac{2\Phi}{c^2}\right)\left(m^2c^2+2m\pder{W}{t}\right)+
\frac{2A^i}{c}\sqrt{m^2c^2+2m\pder{W}{t}}-\\-\sprd{\nabla W}{\nabla
W}=m^2c^2.\end{eqnarray*}Removing the parentheses and omitting
negligibly small terms yields:\be \frac{1}{2m}\sprd{\nabla W}{\nabla
W}-A^i\pder{W}{x^i}+m\Phi= \pder{W}{t}\label{HJshort}\ee that almost
coincides with well-known form of of non-relativistic
Hamilton-Jacobi equation for a particle moving in gravitational
potential $\Phi$ and vector potential $\vec A$. The only difference
is that this equation does not contain the term proportional to
$|\vec A|^2$ which usually presents in the standard version where
1-form $A\equiv A_idx^i$ enters only in combination $\nabla W-A$.
The origin of this difference will be discussed later.

Reduction of the Hamilton-Jacobi equation for isotropic geodesics is
similar and leads to that for geodesics on the 3-space with metric
$g^{ij}$.  Indeed, in this case $m=0$ and all terms containing
negative powers of $c$ can be neglected. Consequently, world lines
of massless particles are just geodesics of the space which they
should be in a non-relativistic theory. As for strictly space-like
geodesics ($m^2<0$), they have no physical meaning in the case.

Usually, when deriving Hamilton-Jacobi equation one starts with
Lagrangian, then introduces generalized momenta, obtains the form of
Hamiltonian and substitutes $\nabla S$ for the momentum. Vector
potential can be included into this scheme two different ways. By
the result, the final form of Hamilton-Jacobi equation depends on
the way it is done. Consider two examples.

Let $\vec u$ be a field of velocities which specifies motion of a
frame w.r.t. some rest or inertial one. Let a mass point move in
this frame without any force acting on it. Then its Lagrangian
$L(\vec v)$ is$$L(\vec v)=\frac{m}{2}(\vec v-\vec
u)^2=\frac{m}{2}{\vec v}^2- m\vec u\cdot\vec v+\frac{m}{2}{\vec
u}^2$$  where the vector $\vec u$ enters as a vector potential and
its square does as scalar potential.

By definition, generalized momentum $\pder{L}{\vec v}$ is$$\vec p=
m(\vec v-\vec u),$$ thus, $$H=\vec p\cdot\vec v-L=\vec p\cdot
\left(\frac{\vec p}{m}+\vec u\right)-\frac{{\vec p}^2}{m}=
\frac{{\vec p}^2}{2m}+\vec p\cdot\vec u,$$ so, square of the vector
potential does not appear in the Hamiltonian and finally, the
Hamilton-Jacobi equation does not contain it. The following example
shows that usually it is not so.

Lagrangian of a particle with unit charge in a vector potential of
magnetostatic field $\vec A$ has the form$$L=\frac{m{\vec v}^2}{2}-
\vec A\cdot\vec v.$$ By definition, generalized momentum is$$\vec p=
m\vec v-\vec A$$ that coincides with the previous case. However,
Hamiltonian is different:$$H=\frac{1}{m}\vec p\cdot(\vec p+\vec A)-L
=\frac{1}{2m}({\vec p}^2+2\vec A\cdot\vec p+{\vec A}^2).$$ This
expression contains square of vector potential that is very
inconvenient when constructing analytical solutions. Thus, if vector
potential specifies motion of the frame chosen, its square appears
in Lagrangian but does not appear in the Hamiltonian and
Hamilton-Jacobi equation whereas if it specifies a kind of
connection its square does not appear in Lagrangian but does in
Hamiltonian and Hamilton-Jacobi equation. Therefore hereafter we
distinguish the two kinds of vector potentials and call that of
gravimagnetic field ``Coriolis potential'' as it was called from the
very beginning by Einstein.

The following remark must be made about rotating frames. If the space
and gravitational field possess axial symmetry it is convenient to use
a coordinate system with one of coordinates being azimuthal angle
$\varphi$ such that $\prt_\varphi$ is the Killing vector. Then passage
to rotating frame is (\ref{Aito}) that produces Coriolis potential \be
A^\varphi\rightarrow A^\varphi+c\omega\left(1+\frac{2\Phi}{c^2}\right)
\label{Arot}\ee.\section{Orthonormal frames and connection}

When reducing Einstein equation we shall employ orthonormal vector
$\{\vec n_a\}$ and co-vector $\{\nu^a\}$ frames and E.~Cartan
structure equations\begin{eqnarray} d\nu^a+\omega_b{}^a\wedge
\nu^b=0 \label{Cartan}\\ \nonumber\Omega_a{}^b=d\omega_a{}^b+
\omega_c{}^b\wedge\omega_a{}^c.\end{eqnarray} It is convenient to
introduce orthonormal frames adapted to the surfaces
$t=const$:\begin{eqnarray}\nu^0=\left(1+\frac{\Phi}{c^2} \right)dt,\
\nu^a=h^a_idx^i-\frac{A^i}{c}dt\, A^a\equiv h^a_iA^i \label{gonfr}\\
\nonumber\vec n_0=\left(1-\frac{\Phi}{c^2}\right) \frac{\prt}{\prt
t}-\frac{A^i}{c}\frac{\prt}{\prt x^i},\ \vec
n_a=h^i_a\frac{\prt}{\prt x^i},\end{eqnarray} where $h_a^i$, $h_i^a$
are matrices obeying the following conditions:$$
\delta^{ab}h_a^ih_b^j=g^{ij},\, h_a^ih_i^b=\delta_a^b.$$ By
construction, the triplet of $\{\nu^a\},\ a\neq0$ constitutes an
orthonormal frame for surfaces $t=const$ because they are orthogonal
to $dt$ and triplet of the vectors $\{\vec n_a\}$ -- an orthonormal
frame for the surfaces because $\vec n_a$ are tangent to them. Since
the frames are adapted to the absolute space as it was defined above
and the space is unaffected by change of frame of reference the
frames (\ref{moframe}) are invariant with respect to transformations
(\ref{Aito}).

Now our task is to obtain the connection 1-form for the frames from
the first structure equation (\ref{Cartan}). Before doing this we
introduce the following triplet of auxiliary 1-forms:$$
\theta^a\equiv h_i^adx^i=\nu^a+\frac{A^a}{c}dt$$ which are not
purely spatial. Exterior derivatives of the 1-forms $\nu^0$ and
$\nu^a$ are:\begin{eqnarray*} d\nu^0=-\frac{1}{c^2}dt\wedge
d\Phi\approx -\frac{1}{c^2}\nu^0\wedge d\Phi\\ d\nu^a=d\theta^a+
\frac{1}{c}dt\wedge dA^a= -\psi_b{}^a\wedge\theta^b+ \frac{1}{c}dt
\wedge dA^a\end{eqnarray*} where we have introduced the connection
1-form $\psi_b^a$ assuming it to satisfy the equation$$d\theta^a+
\psi_b{}^a\wedge\theta^b=0.$$ Thus, $\psi_b{}^a$ is the purely
spatial part of the connection. Indeed,\begin{eqnarray*}d\nu^a=
-\psi_b{}^a\wedge\left(\nu^b+\frac{A^b}{c}dt\right)+
\frac{1}{c}dt\wedge dA^a=\\=-\psi_b{}^a\wedge\nu^b+\frac{1}{c}
dt\wedge\left(dA^a+\psi_b{}^aA^b\right)=-\psi_b{}^a\wedge\nu^b+
\frac{1}{c}dt\wedge DA^a\end{eqnarray*}where $DA^a$ stands for
covariant exterior derivative of the 0-form $A^a$. Finally,$$d\nu^a=
-\psi_b{}^a\wedge\nu^b+\frac{1}{c}\nu^0\wedge DA^a$$ where terms of
order $c^{-3}$ have been ignored.

Solution of the first structure equation for the space-time is
\begin{eqnarray*}\omega_b{}^0=\frac{1}{c^2}(\vec n_b\circ\Phi)\nu^0-
\frac{1}{2c}(D_bA_a+D_aA_b)\nu^a\\ \omega_a{}^b=\psi_a{}^b+
\frac{1}{2c}(D_cA_a-D_aA_c)\delta^{bc}\nu^0 \end{eqnarray*} where
$\vec n_b\circ\Phi$ is action of the differential operator $\vec
n_b$ (\ref{gonfr}) on the function $\Phi$. One more difference
between Coriolis potential $A$ appeared here and genuine vector
potential reveals in these equations: genuine vector potential never
appears in symmetrized derivatives like the last term in the first
line. Hereafter we pass from general case to the case of stationary
axially-symmetric space-time.\section{Stationary axially-symmetric
field}

Let $\{u,v,\varphi\}$ be a coordinate system for axially-symmetric
space in which $\prt_\varphi$ is the Killing vector. All functions
of the field depend only on the coordinates $u$ and $v$. The adapted
frames (\ref{gonfr})
are\begin{eqnarray*}\nu^0=\left(1+\frac{\Phi}{c^2}
\right)dt,\quad\nu^1=\frac{du}{\sigma},\quad\nu^2=\frac{dv}{\sigma},
\quad\nu^3=\rho\left(d\varphi-\frac{A}{c}dt\right)\\
\vec n_0=\left(1-\frac{\Phi}{c^2}\right)\frac{\prt}{\prt t}-
\frac{A}{c}\frac{\prt}{\prt\varphi},\quad\vec n_1=\sigma
\frac{\prt}{\prt u}\quad \vec n_2=\sigma\frac{\prt}{\prt v}\quad\vec
n_3=\frac{1}{\rho}\frac{\prt}{\prt\varphi}.\end{eqnarray*} Note that
in our approximation scheme we can write$$ d\varphi=
\frac{\nu^3}{\rho}+\frac{A}{c}\nu^0.$$ It is easy to obtain the
purely spatial part of connection:$$\psi_1{}^2=
\sigma_v\nu^1-\sigma_u\nu^2=-\psi_2{}^1 \quad\psi_2{}^3=
\frac{\rho_v}{\rho}\sigma\nu^3=-\psi_3{}^2,\quad
\psi_3{}^1=-\frac{\rho_u}{\rho}\sigma\nu^3=-\psi_1{}^3.$$ Below we
use denoting $$ H_{a3}=\vec n_a\circ(\rho A)$$ and keep in mind that
lifting a spatial index changes the sign. Exterior derivatives of
the 1-forms $\nu^a$ are\begin{eqnarray*} d\nu^0&=&-\frac{\vec
n_1\circ\Phi}{c^2}\nu^0\wedge\nu^1- \frac{\vec
n_2\circ\Phi}{c^2}\nu^0\wedge\nu^2,\quad d\nu^1=
-\psi_2{}^1\wedge\nu^2,\quad d\nu^2=-\psi_1{}^2\wedge\nu^1,
\\ d\nu^3&=&\frac{H_1{}^3}{c}\nu^0\wedge\nu^1+
\frac{H_2{}^3}{c}\nu^0\wedge\nu^2-\psi_1{}^3\wedge\nu^1-
\psi_2{}^3\wedge\nu^2.\end{eqnarray*} The corresponding form of
connection is:\begin{eqnarray*} \omega_0{}^1&=&\frac{\vec
n_1\circ\Phi}{c^2}\nu^0+ \frac{H^1{}_3}{2c}\nu^3=
\omega_1{}^0,\quad\quad\omega_1{}^2=\psi_1{}^2 =-\omega_2{}^1\\
\omega_0{}^2&=&\frac{\vec n_2\circ\Phi}{c^2}\nu^0+
\frac{H^2{}_3}{2c} \nu^3=\omega_2{}^0,\quad\quad \omega_2{}^3=
\psi_2{}^3- \frac{H_2{}^3}{2c}\nu^0=-\omega_3{}^2\\ \omega_0{}^3&=&
\frac{H^1{}_3}{2c}\nu^1+\frac{H^2{}_3}{2c}\nu^2=\omega_3{}^0,\quad
\quad\omega_3{}^1=\psi_3{}^1-\frac{H^1{}_3}{2c}\nu^0=-\omega_3{}^2.
\end{eqnarray*} Our next task is to obtain the curvature 2-form of the
space-time that will be done with use of the structure equations
(\ref{Cartan}).\section{Curvature of stationary axially-symmetric
space-time}

Collecting similar terms in the r.h.s. of the second structure
equation (\ref{Cartan}) where exterior derivatives and exterior
products of components of the connection 1-form are substituted, we
obtain the curvature 2-form and the following non-zero components of
the Riemann tensor which contribute the Einstein
equation:\begin{eqnarray}R_0{}^1{}_{01}= -2\frac{D_1(\vec
n_1\circ\Phi)}{c^2}+\frac{H_1{}^3H^1{}_3}{c^2}, \quad\quad
R_0{}^1{}_{02}=-2\frac{D_1(\vec n_2\circ\Phi)}{c^2}+
\frac{H^1{}_3H_{23}}{c^2},\label{4Riem}\\ \nonumber R_0{}^1{}_{31}=
\frac{1}{c}(D_1H^1{}_3+D_3H^3{}_3+D_3H^1{}_1),\quad\quad
R_0{}^2{}_{02}=
-2\frac{D_2(\vec n_2\circ\Phi)}{c^2}+\frac{H_2{}^3H^2{}_3}{c^2},\\
\nonumber R_0{}^2{}_{23}=-\frac{1}{c}
(D_2H^2{}_3+D_3H^3{}_3+D_3H^2{}_2),\quad\quad R_0{}^3{}_{03}=-
2\frac{D_3(\vec n_3\circ\Phi)}{c^2}-\frac{H^2}{2c^2},\\ \nonumber
R_1{}^2{}_{12}=K_1{}^2{}_{12}\quad\quad
R_2{}^3_{}{23}=K_2{}^3{}_{23}- \frac{H_2{}^3H_{23}}{2c^2},\\
\nonumber R_2{}^3{}_{31}=
K_2{}^3{}_{31}+\frac{H_1{}^3H_{23}}{2c^2},\,
R_3{}^1{}_{31}=K_3{}^1{}_{31}-\frac{H_1{}^3H_{13}}{2c^2}.
\end{eqnarray}; $K_a{}^b{}_{cd}$ denotes Riemann tensor of the
space:\be\frac{1}{2}K_a{}^b{}_{cd}\nu^c\wedge\nu^d\equiv
d\psi_a{}^b+\psi_c{}^b\wedge\psi_a{}^c\label{3Riem}.\ee Our next
task is to reduce the Einstein equation to its three-dimensional
form.\section{Ricci and Einstein tensors}

Converting the Riemann tensor (\ref{4Riem}) yields the following
components of the Ricci tensor and scalar
curvature:\begin{eqnarray*} R_{00}&=&-2\frac{\Delta\Phi}{c^2}
+\frac{H^2}{2c^2},\quad\quad R_{03}=\frac{D_aH^a{}_3}{c}\\
R_{11}&=&-2\frac{D_1(\vec n_1\circ\Phi)}{c^2}+
\frac{H_1{}^3H_{13}}{2c^2}+K_{11},\quad R_{12}=-2\frac{D_1(\vec
n_2\circ\Phi)}{c^2}- \frac{H_1{}^3H_{23}}{2c^2}+K_{12},\\
R_{22}&=&-2\frac{D_2(\vec n_2\circ\Phi)}{c^2}+
\frac{H_2{}^3H_{23}}{2c^2}+ K_{22},\quad\quad
R_{33}=-2\frac{D_3(\vec n_3\circ\Phi)}{c^2}+
\frac{H^2}{2c^2}+K_{33}\end{eqnarray*} where $\Delta\Phi$ is
commonplace three-dimensional Laplacian$$\Delta\Phi=
\delta^{ab}D_a(\vec n_b\circ\Phi)$$ and we put $D_3H^a{}_a=0$
because trace of $H_{ab}$ is zero; $K_{ab}$ and $K$ -- Ricci tensor
of the space:$$ K_{ab}=K_a{}^c{}_{bc}$$ and its trace. Scalar
curvature of the space-time is$$ R= -\frac{4}{c^2}-K.$$ As expected,
this scalar does not depend on $H_a{}^b$ and, hence, on choice of
the reference frame. Now we can compose the Einstein tensor. Its
components are:\begin{eqnarray*}
G_{00}&=&\half\left(K+\frac{H^2}{c^2}\right),\ G_{03}=
\frac{1}{c}D_aH^a{}_3,\\ G_{11}&=&\frac{2}{c^2} [D_1(\vec
n_1\circ\Phi)-\Delta\Phi]+\Gamma_{11}+ \frac{H_1{}^3H_{13}}{2c^2}\\
G_{12}&=&\frac{2}{c^2}D_1(\vec n_2\circ\Phi)+\Gamma_{12}-
\frac{H_1{}^3H_{23}}{2c^2}\\ G_{22}&=&\frac{2}{c^2} [D_2(\vec
n_2\circ\Phi)-\Delta\Phi]+\Gamma_{22}+ \frac{H_2{}^3H_{23}}{2c^2}\\
G_{33}&=&\frac{2}{c^2} [D_3(\vec n_3\circ\Phi)-\Delta\Phi]+
\Gamma_{33}+\frac{H^2}{c^2}\end{eqnarray*} where we have introduced
Einstein tensor for the space:$$ \Gamma_{ab}=K_{ab}-
\half\delta_{ab}K.$$ Finally, Einstein equation reduces to the
following system:\begin{eqnarray*}
\half\left(K+\frac{H^2}{c^2}\right)=\kappa T_{00}=
\kappa\varepsilon,\quad\quad\frac{1}{c}D_aH^a{}_3=\kappa
T_{03}=\kappa J_3\label{EE}\\ \frac{2}{c^2}[D_1(\vec n_1\circ\Phi)-
\Delta\Phi]+\Gamma_{11}+\frac{H_1{}^3H_{13}}{2c^2}=\kappa T_{11}\\
\frac{2}{c^2}D_1(\vec n_2\circ\Phi)+\Gamma_{12}-
\frac{H_1{}^3H_{23}}{2c^2}=\kappa T_{12}\\ \frac{2}{c^2} [D_2(\vec
n_2\circ\Phi)-\Delta\Phi]+\Gamma_{22}+ \frac{H_2{}^3H_{23}}{2c^2}
=\kappa T_{22}\\ \frac{2}{c^2} [D_3(\vec n_3\circ\Phi)-
\Delta\Phi]+\Gamma_{33}+\frac{H^2}{c^2}=\kappa T_{33}\end{eqnarray*}
where $T_{ab}$ stands for stress-energy tensor of the matter; its
components $\varepsilon$ and $J_3$ are density of energy (including
rest energy $\rho c^2$) and the 3-component of the mass current.
Remarkably, one of these equations reads$$\Delta\Phi+\quart K=4\pi
k\mu,\ k=\frac{16\pi}{c^2}$$ where $k$ is Newtonian gravitational
constant, $\mu$ -- mass density.\section{Acknowledgments}

This work was supported in part by TWAS (The Academy of Sciences for
the Developing World) Associateship program and completed in the
Institute for Advanced Studies in Basic Sciences (IASBS) Zanjan,
Iran. The author would like to express his gratitude to Dr. Y.
Sobouti and Dr. H. Safari for critical discussions and Dr. M.
Esfahani-Zadeh for facilitating the communication of the
mathematical content of this work.

\end{document}